\documentclass[]{aa}
\usepackage{graphicx}
\usepackage{amsmath}
\usepackage{tablefootnote}
\usepackage{bm}
\usepackage{multirow}
\usepackage[varg]{txfonts}
\usepackage{float}

\usepackage[colorlinks,linkcolor=blue,citecolor=blue,linktocpage=true,breaklinks, 
plainpages=false,urlcolor=blue]{hyperref}

\bibpunct{(}{)}{;}{a}{}{,}
\begin{document}

\title{Probing photospheric magnetic fields with new spectral line pairs}
\titlerunning{Probing photospheric magnetic fields with new spectral line pairs}
\authorrunning{Smitha \& Solanki}

\author{H. N. Smitha \inst{1} \and S. K. Solanki \inst{1,2}}
\institute{$^{1}$Max-Planck-Institut f\"ur Sonnensystemforschung, Justus-von-Liebig-Weg 3, 
37077 G\"ottingen, Germany\\
$^{2}$School of Space Research, Kyung Hee University, Yongin, Gyeonggi, 446-701, Republic 
of Korea\\
\email{smitha@mps.mpg.de}}
\abstract
% Context
{The magnetic line ratio (MLR) method has been extensively used in the measurement of 
photospheric magnetic field strength. It was devised for the neutral iron line pair at  
5247.1\,\AA{} and 5250.2\,\AA{} (5250\,\AA{} pair). Other line pairs as well-suited as 
this pair been have not been reported in the literature.}
% Aim
{The aim of the present work is to identify new line pairs useful for the MLR technique 
and to test their reliability.}
% Methods
{We use a three dimensional magnetohydrodynamic (MHD) simulation representing the quiet 
Sun atmosphere to synthesize the Stokes profiles. Then, we apply the MLR technique to the 
Stokes $V$ profiles to recover the fields in the MHD cube both, at original resolution and 
after degrading with a point spread function.  {In both these 
cases, we aim to empirically represent the field strengths 
returned by the MLR method in terms of the field strengths in the MHD 
cube.}}
% Results
{We have identified two new line pairs that are very well adapted to be used for MLR 
measurements. The first pair is in the visible, Fe~{\sc i} 6820\,\AA--6842\,\AA{} (whose 
intensity profiles have earlier been used to measure stellar magnetic fields), and the 
other is in the infrared (IR), Fe~{\sc i} 15534\,\AA--15542\,\AA.  The lines in these 
pairs reproduce the magnetic fields in the MHD cube rather well, partially better than the 
original 5250\,\AA{} pair.}
% Conclusions
{The newly identified line pairs complement the old pairs. The lines in the new IR pair, 
due to their higher Zeeman sensitivity, are ideal for the measurement of weak fields. The 
new visible pair works best above $300$\,G. The new IR pair, due to its large Stokes $V$ 
signal samples more fields in the MHD cube than the old IR pair at $1.56\,\mu$m, even in 
the presence of noise, and hence likely also on the real Sun. Owing to their low formation 
heights (100--200\,km above $\tau_{5000}=1$), both the new line pairs are well suited for 
probing magnetic fields in the lower photosphere.}
 
\maketitle

\section{Introduction}
Spectral lines offer diagnostics to measure magnetic fields on the Sun. Accurate 
magnetic field measurement relies on an optimal combination of spectral lines and the 
method employed to extract the information on the field. Unfortunately, the Stokes 
profiles $(I,Q,U,V)$ are affected by many other atmospheric parameters besides the 
field, making the extraction of the field complex and time consuming. To bypass this, 
at least for the field strength, \citet{1973SoPh...32...41S} proposed the magnetic line 
ratio (MLR) method which involves determining the intrinsic magnetic field strength ($B$) 
from the ratio of Stokes $V$ of two lines. The two spectral lines must form under the same 
atmospheric conditions but differ in their magnetic sensitivities, given by the effective 
Land\'{e} g-factors (g$_{\rm eff}$). For weak, height-independent fields, the Stokes $V$ 
ratio is simply equal to the ratio of their g$_{\rm eff}$. In the presence of strong 
height-independent fields, the ratio saturates and becomes independent of $B$. This method 
works best for intermediate field strengths where the Stokes $V$ ratio is proportional to 
$B$, due to the differential Zeeman saturation. \citet{1973SoPh...32...41S} applied MLR 
to the line pair Fe~{\sc i} 5247.1\,\AA--5250.2\,\AA{} (5250\,\AA{} pair) in the 
photospheric network which lead to the discovery of the presence of kilo-Gauss (kG) 
fields. Since then the MLR has been widely used to measure photospheric magnetic fields 
\citep{1985SoPh...95...99S, 1987A&A...188..183S, 1988A&A...192..338S, 1992A&A...263..312S, 
1994A&A...286..626K, 1998A&A...337..928G, 1999ESASP.448..853L, 2010A&A...517A..37S, 
2011A&A...529A..42S}. For reviews on MLR see \citet{1993SSRv...63....1S, 
2009ASPC..405..135S, 2009SSRv..144..275D, 2013A&ARv..21...66S}.

In addition to the MLR, line pairs formed at similar heights in the atmosphere but 
with different g$_{\rm eff}$ are used in multi-line inversions to measure 
magnetic field. Two other spectral line pairs, used for inversions as well as MLR, 
are the 6301.5\,\AA--6302.5\,\AA{} (6300\,\AA{} pair) in the visible 
\citep[][]{2003ApJ...582L..55D, 2003A&A...407..741D, 2010A&A...517A..37S, 
2011ApJ...735...74I, 2012ASPC..456....3S}, and the 15648.5\,\AA--15652.8\,\AA{} 
($1.56\,\mu$m pair) in the infrared \citep[IR, e.g.,][]{1996A&A...310L..33S}. The 
$1.56\,\mu$m pair was identified by \citet{1992A&A...263..312S} and has been used in the 
measurement of internetwork fields \citep{1995ApJ...446..421L, 1996A&A...310L..33S, 
2003A&A...408.1115K, 2007A&A...469L..39M, 2016A&A...596A...6L}. The 6300\,\AA{} pair is 
used to study both quiet and active regions on the Sun \citep[for 
e.g.,][]{2003ApJ...582L..55D, 2003A&A...407..741D, 2002ApJ...565.1323S, 
2003ApJ...593..581S, 2004ApJ...611.1139S, 2006A&A...456.1159M, 2007ApJ...666L.137C, 
2008ApJ...672.1237L}, based on the observations from both ground based telescopes and the 
\textit{Hinode} satellite. The distribution of the quiet Sun photospheric magnetic field 
revealed by these two line pairs is quite different, particularly in the internetwork, 
first observed by \citet{2003ApJ...597L.177S}. The $1.56\,\mu$m pair indicates the 
presence of mostly sub-kG fields while the 6300\,\AA{} pair indicates kG fields. Combined 
analyses of the two line pairs have been carried out by \citet{2003ApJ...593..581S, 
2005A&A...442.1059K, 2006ApJ...646.1421D}, although with contradictory results.

\begin{table*}
\label{table1}
\centering
\caption{Atomic parameters of the new and the old line pairs.}
\begin{tabular}{cccccccccc}
\hline
& & & & & & & & & \\
\multicolumn{10}{c}{\textbf{New pairs}}\\
& & & & & & & & & \\
\hline
&Wavelength\,(\AA)& Element &  $J_l$ & $J_u$ &log($gf$) & 
$\chi_{e}$\,(e.v.) & 
g$_l$ & 
g$_u$ & g$_{\rm eff}$ \\
\hline
\multirow{2}{*}{I}&6820.3715 & Fe~{\sc i} &  1.0 & 2.0 & -1.32 & 4.63 & 2.5 & 1.83 
&1.5 \\	
&6842.6854 & Fe~{\sc i} &  1.0 &1.0 & -1.32 & 4.63 & 2.5 & 2.5 &2.5 \\

\multirow{2}{*}{II}&6213.4291 & Fe~{\sc i} &  1.0 &1.0 &-2.48 & 2.22 & 2.5 & 1.5 & 
2.0 \\
&6219.2802 & Fe~{\sc i} &  2.0 &2.0 &-2.43 & 2.19 & 1.83 & 1.5 & 1.67 \\

\multirow{2}{*}{III}&15534.257 & Fe~{\sc i} &  1.0 &2.0 &-0.382 & 5.64 & 1.5 & 1.83 
& 2.0 \\
&15542.089 & Fe~{\sc i} &  1.0 &0.0 &-0.337 & 5.64 & 1.5 & 0.0 & 1.50 \\
\hline
& & & & & & & & & \\
\multicolumn{10}{c}{\textbf{Old pairs}}\\
& & & & & & & & & \\
\hline
 &Wavelength\,(\AA)& Element & $J_l$ & $J_u$ &log($gf$) & $\chi_{e}$\,(e.v.) 
& g$_l$ & 
g$_u$ & g$_{\rm eff}$ \\
\hline
\multirow{2}{*}{I}&5247.0504 & Fe~{\sc i} & 2.0 &3.0&-4.946 & 0.087 & 1.5 & 1.75 &2.0\\
&5250.2080 & Fe~{\sc i} &  0.0 &1.0 &-4.938 & 0.121 & 0.0 & 3.0 &3.0\\

\multirow{2}{*}{II}&6301.5012 & Fe~{\sc i} & 2.0 & 2.0 & -0.718 & 3.654 & 1.83 & 
1.5 &1.67\\
&6302.4936 & Fe~{\sc i} & 1.0 & 0.0 & -1.236 & 3.686 & 2.5 & 0.0 &2.5\\

\multirow{2}{*}{III}&15648.518 & Fe~{\sc i} &  1.0 & 1.0 & -0.675 & 5.426 & 
3.0 &3.0 &3.0\\
&15652.874 & Fe~{\sc i} & 5.0 & 4.0 & -0.043 & 6.246 & 1.51&1.49&1.53\\

\multirow{2}{*}{IV}&4122.8020 & Fe~{\sc i} &  2.0 & 3.0 & -1.300 & 2.832 & 
1.50 &1.16 &0.820\\
&8999.5600 & Fe~{\sc i} & 2.0 & 2.0 & -1.300 & 2.832 & 1.50&1.49&1.496\\
\hline
\end{tabular}
\tablefoot{The columns indicate 
wavelength, ion, multiplet number, total angular momentum quantum number of the lower 
($J_l$) and upper levels ($J_u$), the oscillator strength log$(gf)$, the lower level 
excitation potential in e.v. ($\chi_e$), Land\'{e} g-factors of the lower (g$_l$) and 
upper levels (g$_u$), and the effective Land\'{e} g-factor (g$_{\rm eff}$), respectively.}

\end{table*}

\citet{2007ApJ...659.1726K} applied MLR to the above three line pairs synthesized from a 
three dimensional magnetohydrodynamic (3D MHD) snapshot. They concluded in favour of the 
$1.56\,\mu$m  and the 5250\,\AA{} pairs, although they could not recover kG from the 
5250\,\AA{} pair. The 6300\,\AA{} pair does not reproduce the fields in the MHD cube, due 
to the difference in height of formation (HOF) of the two lines 
\citep[][]{2001ApJ...550..970S, 2007ApJ...659.1726K, 2010A&A...514A..91G}. 
Discrepancies in the results from the 6300\,\AA{} pair observations have been reported in 
\citet{2003A&A...407..741D,2006A&A...456.1159M}. In a contrasting study, 
\citet{2008ApJ...674..596S}, concluded that the 6300\,\AA{} pair is better than the 
5250\,\AA{} pair, as they could not recover kG fields in the network observations from 
the 5250\,\AA{} pair. In Section~\ref{spdg-section} of the present paper, we try to 
provide an explanation for this discrepancy. In order to compensate for the 
difference in HOF of the 6300\,\AA{} pair, \citet{2010A&A...517A..37S, 
2013A&A...556A.113S} devised a renormalization to the MLR of 6300\,\AA{} pair in terms of 
the 5250\,\AA{} pair. 

Any differences in the HOF of the lines in a pair increase the difficulties in 
interpreting the results from the MLR. The formation heights of the lines in 6300\,\AA{} 
pair are separated by more than 100\,km and those in the $1.56\,\mu$m pair by 
$\approx 30$km. These issues leave us with a single ``ideal'' line pair 
\citep{2013A&A...556A.113S} for the MLR. \citet{2007A&A...465..339S, 2008ApJ...674..596S} 
proposed that the pair 4122\,\AA{}--9000\,\AA{} works better than all the above pairs 
(i.e., pairs I, II and III in the lower part of Table~\ref{table1}). However 
these lines are 5000\,\AA{} apart and need to be observed simultaneously. Also, the 
Land\'{e} g-factor of the 4122\,\AA{} is quite low (see Table~\ref{table1}) and hence 
this line is less sensitive to magnetic fields. A survey of the Fe~{\sc i} lines with 
different magnetic sensitivities was carried out by \citet{2009KPCB...25..319V}. They 
present a list of 28 line pairs which are suitable for MLR. However most of them are quite 
weak and the authors do not discuss the reliability of these pairs in 
detail.

After a detailed search in the visible and IR range of the solar 
spectrum, we have identified three new line pairs. Two pairs are in 
the visible at 6820\,\AA--6842\,\AA{} (6842\,\AA{} pair) and 
6213\,\AA-6219\,\AA{} (6219\,\AA{} pair). The third pair is in the IR 
at 15534\,\AA--15542\,\AA{} ($1.55\,\mu$m pair). The lines in each 
pair have identical/similar atomic parameters but different g$_{\rm 
eff}$. We find that the 6842\,\AA{} and the $1.55\,\mu$m pairs are 
more suitable for MLR than the 6219\,\AA{} pair. The lines in these 
two pairs are formed deep in the photosphere.  {We compare 
the performance of the new and the old line pairs by applying the MLR 
method to the Stokes profiles in a 3D MHD cube, and by comparing the 
results with the fields in the cube.  This is done at both 
original resolution of the cube and after applying a degradation. In 
the first case, we show that the magnetic field strengths given by MLR 
are best represented when the field strengths in the MHD cube 
are weighted by the response functions of Stokes $V$ profiles and 
then integrated over the optical depth. In the presence of 
instrumental degradation, this is quite challenging.  In this 
paper, we have made the first attempt to empirically represent the 
magnetic field strengths returned by the MLR method in a realistic 
atmosphere with realistic degradation.}

In Section~\ref{new}, we discuss the atomic parameters of the new 
lines. In Section~\ref{hof}, we compute their HOF from the response 
functions. A detailed comparison between the $B$ from the MLR and the 
MHD cube is presented in Section~\ref{mhd-mlr-section}. In 
Section~\ref{spdg-section}, we repeat the analyses by spatially and 
spectrally degrading the Stokes profiles and present the conclusions 
in Section~\ref{conclusions}.
%===========================================================
\section{Atomic parameters}
\label{new}
The new and the old line pairs are listed in Table~\ref{table1}. The atomic 
parameters are taken from 
Kurucz\footnote{\url{http://kurucz.harvard.edu/linelists.html}}, 
NIST\footnote{\url{http://www.nist.gov/pml/data/asd.cfm}} and 
VALD\footnote{\url{http://vald.astro.uu.se/}} atomic databases. 

The newly identified pairs have been listed in 
\citet{1985A&A...148..123S, 1992A&A...263..312S, 1995A&AS..113...71R}. 
The intensity profiles of the 6842\,\AA{} line pair has earlier been 
used by \citet{1997A&A...318..429R} to measure the magnetic fields on 
cool dwarfs using inversions. Due to its large g$_{\rm eff}$, the 
6842\,\AA{} line is used by \citet[][]{1993A&A...279..243B, 
2000SoPh..197..227W} to study sunspots and filaments. Furthermore, the 
6842\,\AA{} line in combination with Fe {\sc i} 6843\,\AA{} line was 
used by \citet{1994ASPC...64..474S} to measure magnetic fields in 
late-type stars. The two lines in the  6842\,\AA{} are separated by 
22\,\AA{} and are unblended. They have identical oscillator strengths 
(log$(gf)$) and excitation potentials ($\chi_e$) with very different 
g$_{\rm eff}$. In the absence of a magnetic field, the lines are 
formed at the same height in the atmosphere and further details will 
be discussed in Section~\ref{ht-rf}. Due to their high excitation 
potential, these lines are less sensitive to fluctuations in $T$ than 
the 5250\,\AA{} pair. 
 
The lines in the 6219\,\AA{} pair, have the same log($gf$) and nearly 
same $\chi_e$.  Though they are formed at the same height in the 
atmosphere for $B=0$, their g$_{\rm eff}$ differ by only 20\%, 
rendering them non-ideal for MLR. The third new pair is in the IR, 
separated by 8\,\AA{} at 15534\,\AA--15542\,\AA{}. They belong to 
different multiplets but have the same $\chi_e$ and similar log($gf$). 
The red wing of the 15534\,\AA{} line is affected by a minor 
unidentified blend which, may not significantly affect the Stokes $V$ 
profiles. The blue wing is clean without any blends. {The 15542\,\AA{} 
line has no visible blends in the solar spectrum, however, 
\citet{1990A&AS...83..307S} indicate the presence of three Mg~{\sc i} 
blends. These three Mg~{\sc i} lines have not been listed in the 
Kurucz, NIST or VALD atomic databases {and appear to have been present 
only in older databases, so that they may be spurious.} According to 
\citet{1995A&AS..113...71R} this line is only lightly blended by a 
Si~{\sc i} line at 15542.016\,\AA{}.}

The new line pairs in the visible and IR are separated by 22\,\AA{} 
and 8\,\AA{}, respectively. The IR pair can be observed with the 
GREGOR Infrared Spectrograph \citep[GRIS;][]{2012AN....333..872C} as 
spectral range as wide as 20\,\AA{} has been observed with this 
instrument \citep{2016A&A...596A...6L}. It is possible to cover the 
22\,\AA{} range of the visible line pair using 2k x 2k detector at a 
spectral resolving power of 270000 or better. The new line pairs can 
be observed with the spectro-polarimeters at the upcoming Daniel K. 
Inouye Solar Telescope (DKIST) such as the Visible Spectro-Polarimeter 
\citep[ViSP;][]{2014SPIE.9147E..07E} 
and the 
Diffraction Limited Near Infrared Spectro-Polarimter 
\citep[DL-NIRSP;][]{2014SPIE.9147E..07E}.

\begin{figure}
\centering
\includegraphics[width=0.45\textwidth]{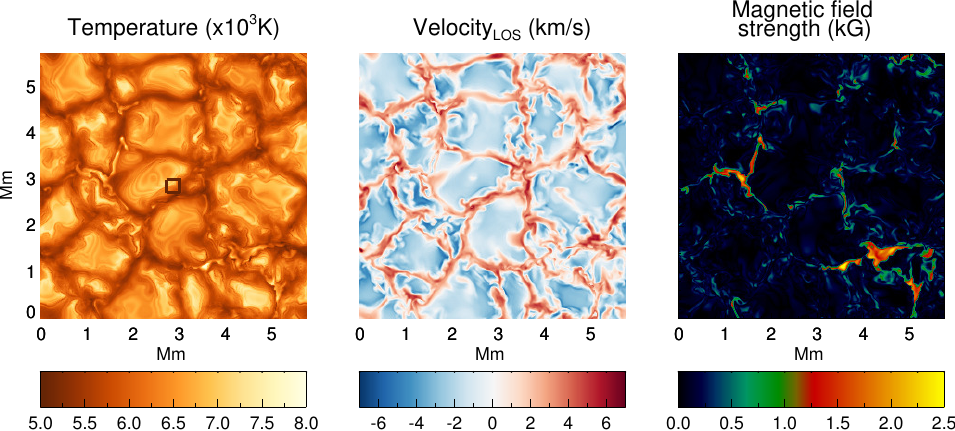}
\caption{The temperature, line-of-sight (LOS) velocity and 
magnetic field maps of the 
MURaM MHD cube at log($\tau_{5000})=0$. The cube has an average unsigned LOS magnetic 
field of 50\,G.  {The black box near the center of the first panel 
represents the area over which 
the Stokes profiles in Figure~\ref{synth_spectra} are averaged. The size of the box is 
$0.4^{\prime\prime} \times 0.4^{\prime\prime}$.} }
\label{muram-prop1}
\end{figure}

\begin{figure*}
\centering
\includegraphics[width=0.8\textwidth]{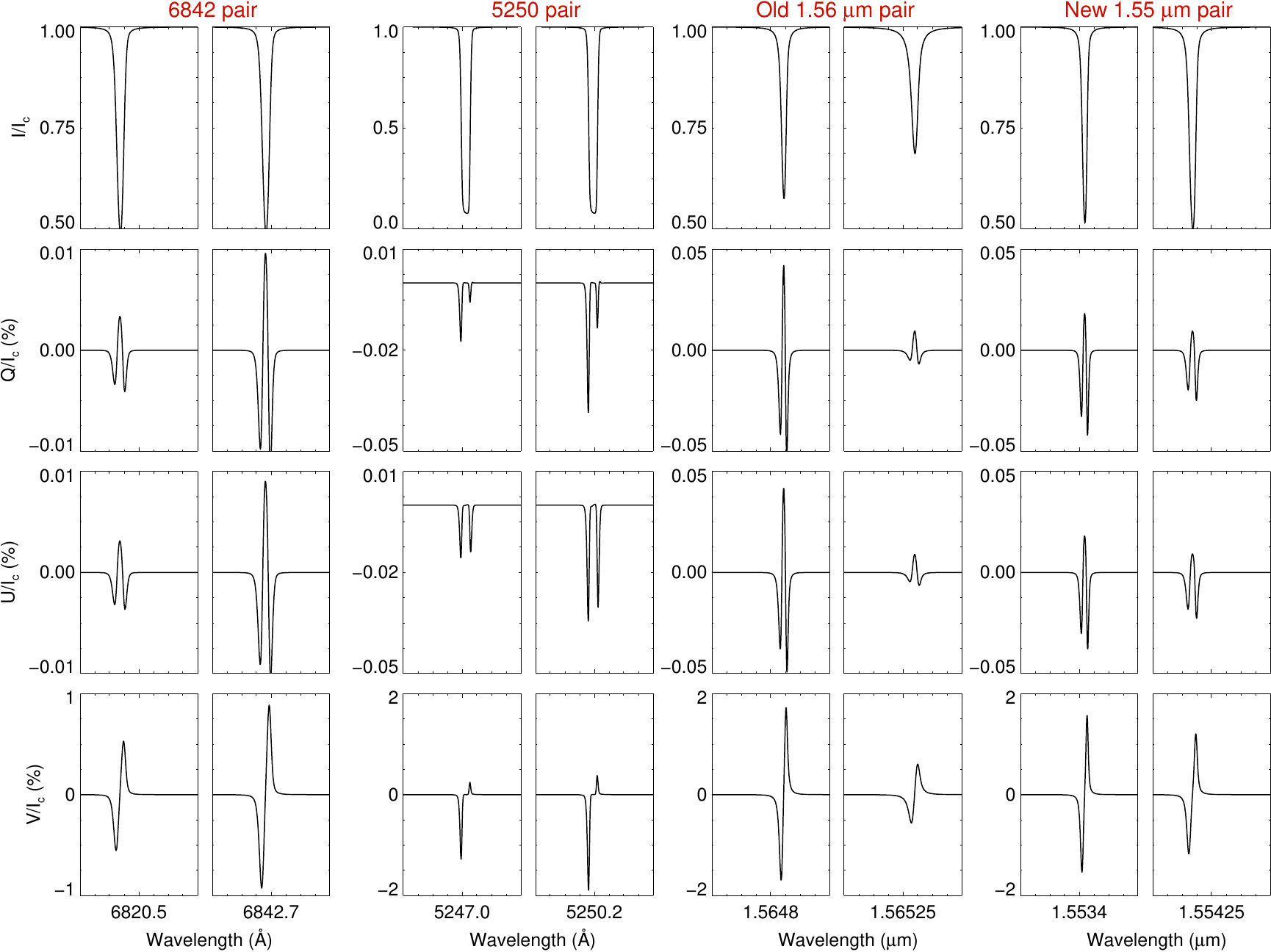}
\caption{ {Stokes profiles of the four line pairs from the MHD 
cube, averaged over a box of size $0.4^{\prime\prime} \times 0.4^{\prime\prime}$. The 
location of the box is shown in the first panel of Figure~\ref{muram-prop1}.}}
\label{synth_spectra}
\end{figure*}

\section{Height of formation}
\label{hof}
\subsection{3D MHD cube and profile synthesis}
\label{muram-cube}
We use a snapshot of a 3D MHD simulation computed from the MURaM code 
\citep{2005A&A...429..335V}. We have selected a cube from 
the set used by \cite{2014A&A...568A..13R} but with a different resolution. The size of 
the cube is (6 $\times$ 6 $\times$ 1.4)\,Mm with a resolution of (20.83 $\times$ 20.83 
$\times$ 14)\,km. The cube has an unsigned average line-of-sight (LOS) magnetic field of 
50~G. The properties of the cube, such as the temperature ($T$), LOS 
velocity ($\varv_{\rm LOS}$), and $B$ at log($\tau_{5000}$)=0 are 
shown in Figure~\ref{muram-prop1}. The cube represents the quiet Sun 
atmosphere, dominated by weak and intermediate fields. However 
there are a few patches of strong magnetic fields in the lower right 
corner and in the mid-left, as seen from Figure~\ref{muram-prop1}. The 
Stokes profiles are synthesized using the 
Stokes-Profiles-INversion-O-Routines (SPINOR) of 
\citet[][]{2000A&A...358.1109F,phdthesis}, run in its forward mode 
along each vertical column of the MHD cube (1.5D) at a heliocentric 
angle of $\mu=1$.

 {In Figure~\ref{synth_spectra}, we present the Stokes profiles of all 
the four 
line pairs spatially averaged over a small region of $0.4^{\prime\prime} \times 
0.4^{\prime\prime}$ close to the center of the analyzed MHD snapshot, indicated by the black box
in the first panel of Figure~\ref{muram-prop1}. The size of the averaged area is chosen to match the resolution 
of recent observations at the GREGOR telescope \citep{2012AN....333..796S} with the GREGOR Infrared Spectrograph (GRIS)
instrument \citep{2012AN....333..872C} such as those presented in \citet{2016A&A...596A...6L}}. 

In the visible range, the new 6842\,\AA{} pair is weaker than the 5250\,\AA{} pair in 
both intensity and polarization (note the different vertical scales). However, in the IR, 
though the 15648\,\AA{} line of the old IR pair is strong and has large Stokes amplitudes ($Q,U,V$), the 
15652\,\AA{} line has much weaker amplitudes, especially in $Q$ and $U$ \citep[see 
also][]{2008A&A...477..953M,2016A&A...596A...6L} than the lines of the new pair. This
makes it harder to use the lines of the old IR pair together, in MLR 
as well as in inversions when the profiles are affected by noise. In 
this respect, the new $1.55\,\mu$m pair offers great advantage as both 
the lines have large Stokes amplitudes. The strong linear polarization 
signals can be particularly favourable for measuring vector magnetic 
fields using inversions.

\subsection{Response functions}
\label{ht-rf}
\begin{figure}
\begin{center}
\includegraphics[width=0.48\textwidth]{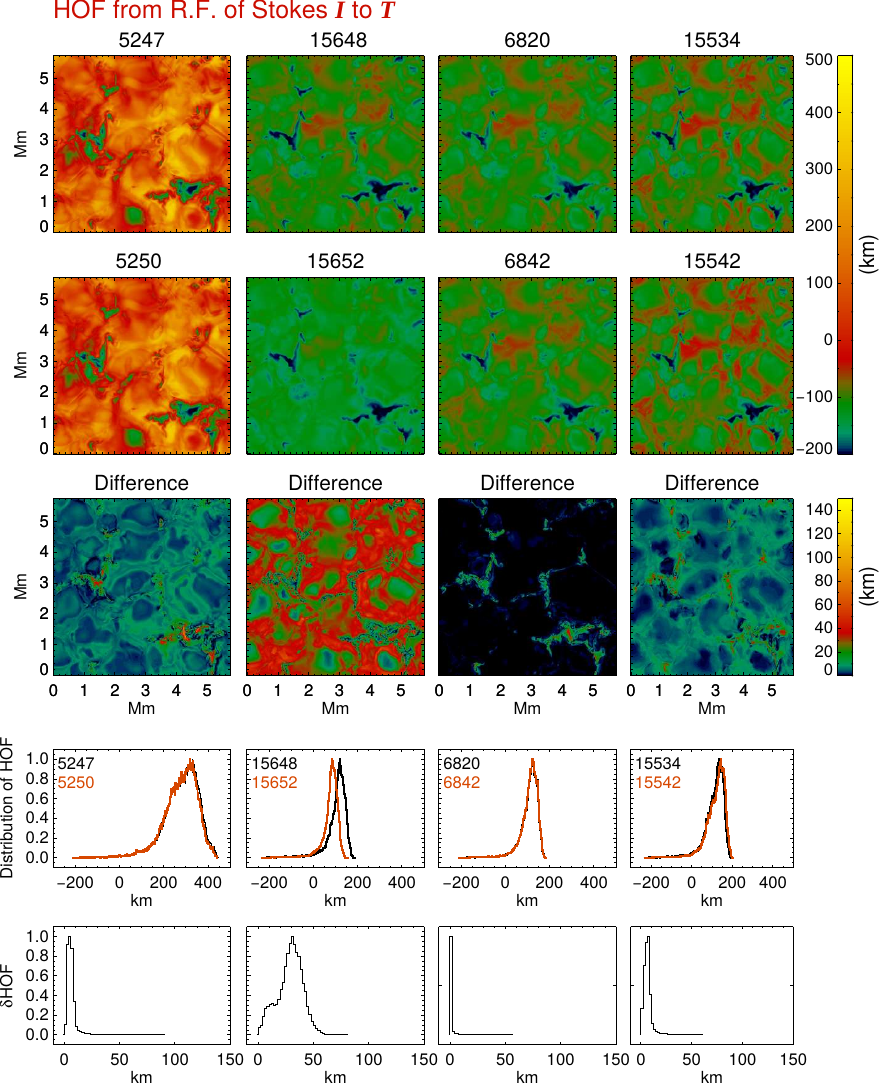}
\caption{ {Maps of height of formation (HOF) for different line 
pairs deduced from the centroid of the response function (RF) of 
Stokes $I$ profiles to temperature, at the central wavelength.} First 
and second rows: distribution of HOF referring to individual spatial 
pixels of the MHD cubes. Third row: difference (absolute) in the HOF. 
Forth row: histogram of the HOF of the two lines in each pair. Fifth 
row: histogram of the absolute differences in the HOF.}
\label{rfs}
\end{center}
\end{figure}

To compare the HOF of the line pairs, we use \textit{Response 
Functions} \citep[RFs,][]{1975SoPh...43..289B}. These functions 
measure the responses of the line profiles to variations in 
atmospheric properties such as $T$, $\varv_{\rm LOS}$ and $B$. 
 {Using the SPINOR code, we compute the RFs of the Stokes $I$ 
and $V$ profiles of all the lines to these three atmospheric 
properties. For the Stokes $I$ profiles, we use the RF at the line 
center wavelength and for Stokes $V$, we use the RF at the 
wavelength corresponding to the largest peak in the $V$ profile. This 
is because later, in section~\ref{mhd-mlr-section}, we compute the 
MLR from this largest peak, also referred to as the prominent peak.} 
The HOF is then assumed to be at the centroid of the RFs. The 
distribution of the HOF across the cube, for different spectral lines, 
 {from RFs of Stokes $I$ and $V$ profiles to $T$ are shown in 
the first two rows of Figures~\ref{rfs} and \ref{rfsv}, respectively}. 
The third row is the unsigned difference in the HOF ($\delta$HOF) 
between the lines in the pair. The histograms of the distribution of 
HOF and 
$\delta$HOF are shown in the last two rows. The reference height 
$z=0$\,km corresponds to the geometrical layer where 
log$(\tau_{5000})$, on average, is zero. This HOF represents the 
atmospheric height that is most sampled by the spectral line. In other 
words, the Stokes profiles are strongly influenced by the physical 
conditions at the HOF of the line.

Though $T$ has a dominant influence on the spectral lines and 
their formation, the RFs from $\varv_{\rm LOS}$, $B$ and magnetic 
field inclination ($\gamma$) also provide valuable information, 
especially for the MLR. Gradients in the $\varv_{\rm LOS}$, $B$ and 
$\gamma$ affect the shapes of the Stokes $V$ profiles, resulting in 
asymmetries \citep{2005A&A...442.1059K}. Hence, the MLR works best if 
the two lines sample the same $\varv_{\rm LOS}$, $B$ and $\gamma$, in 
addition to $T$.  {To confirm this, we have computed 
the $\delta$HOF for each line pair from the RFs of Stokes $I$ and $V$ 
profiles to $\varv_{\rm LOS}$, $B$, and $\gamma$ RFs, in the same way 
as we did for the $T$ RFs. The variations in $\delta$HOF and 
the histograms are shown in Figures~\ref{rfs1} and \ref{rfsv1}.}

\subsubsection{Old pairs}
\begin{figure}
\centering
\includegraphics[width=0.48\textwidth]{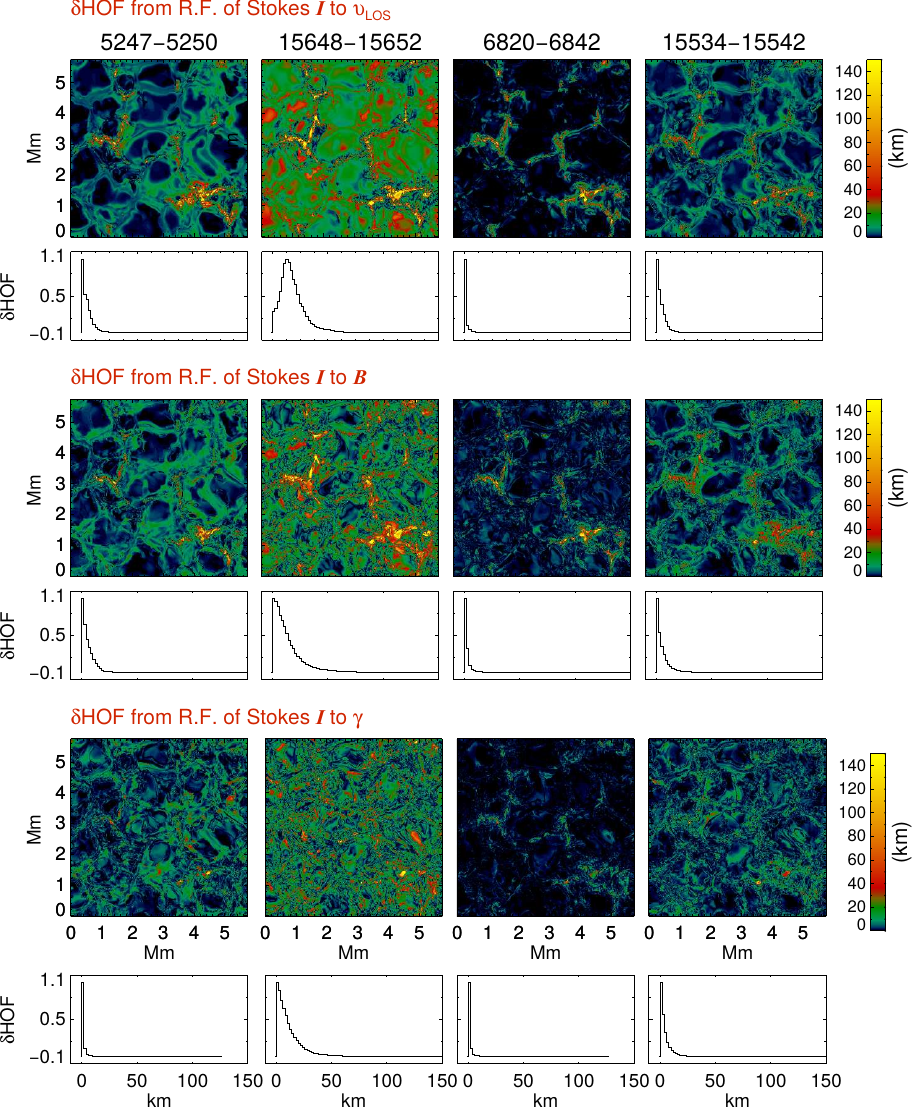}
\caption{ {Maps of difference in heights of formation and their 
histograms. They 
are similar to the third and fifth rows of Figure\,\ref{rfs}, but 
computed from the RFs of Stokes $I$ profiles to perturbations in 
velocity, magnetic field strength, and inclination.}}
\label{rfs1}
\end{figure}

\begin{figure}
\centering
\includegraphics[width=0.48\textwidth]{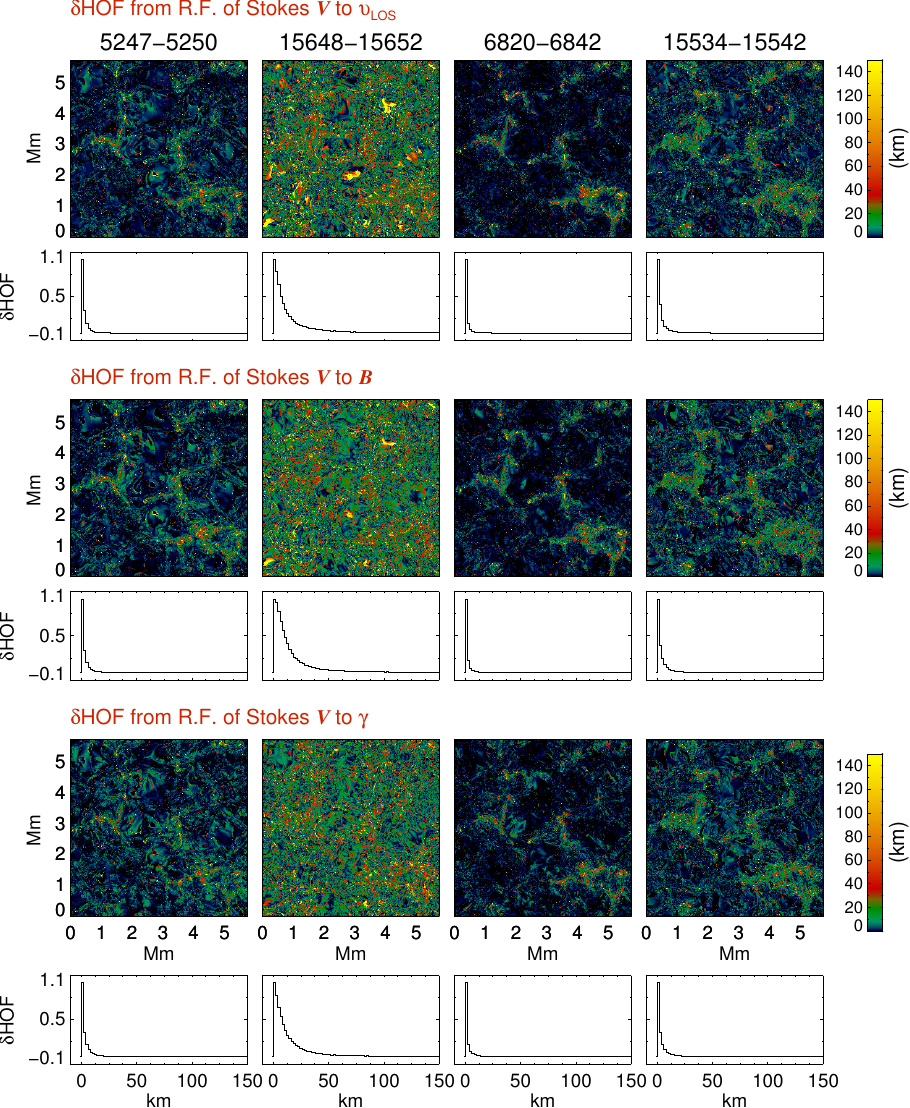}
\caption{ {Maps of difference in heights of formation and their 
histograms, 
similar to Figure~\ref{rfs1}, but computed from the response functions 
of Stokes $V$ profiles to perturbations in velocity, magnetic field 
strength, and inclination.}}
\label{rfsv1}
\end{figure}

From the $T$ RFs of  {Stokes $I$ (Figure~\ref{rfs})}, 
the {5247\,\AA} and {5250\,\AA} lines are formed at similar heights in 
the atmosphere for $B=0$ and for weak fields. The histograms of the 
distribution of HOFs obtained for vertical rays passing through each 
horizontal pixel of the MHD snapshot for both the lines almost 
entirely overlap and peak around 300\,km and the $\delta$HOF has a 
narrow spread with a peak at 20\,km. They start to differ for 
intermediate and strong fields where the $\delta$HOF can be as high as 
100~km. These pixels correspond mostly to the edges of granules to 
intergranular lanes. In regions of strong magnetic field 
concentrations (seen at bottom right and mid-left of the atmosphere), 
not only the HOF of both lines decrease due to plasma evacuation 
leading to a drop in the gas pressure, the $\delta$HOF also increases 
due to the difference in Land\'e factor. The histograms of the 
$\delta$HOF deduced from  $\varv_{\rm LOS}$, $B$ and $\gamma$ RFs 
 {of Stokes $I$ profiles at the line center 
(Figure~\ref{rfs1})}, peak close to zero and at a few pixels reach 
values as high as 150\,km in the strong magnetic regions. Differences 
in HOF are also seen in the regions 
surrounding the strong field concentrations because of the magnetic 
canopies, leading to the measurement of stronger $B$ from MLR 
\citep[Section~\ref{mhd-mlr-section}, see 
also][]{2007ApJ...659.1726K}. 

The $T$ RFs  {of Stokes $I$ profiles (Figure~\ref{rfs})} 
of the 1.56\,$\mu$m pair indicate that the two lines are most commonly 
formed around 30\,km apart. However this difference decreases for the 
$\varv_{\rm LOS}$, $B$ and $\gamma$ RFs  {(Figure~\ref{rfs1})}. 
In particular, the two lines sample similar $B$, despite the 
difference in their HOFs  {from the $T$ RFs of Stokes 
$I$ profiles}.

\begin{figure}
\begin{center}
\includegraphics[width=0.48\textwidth]{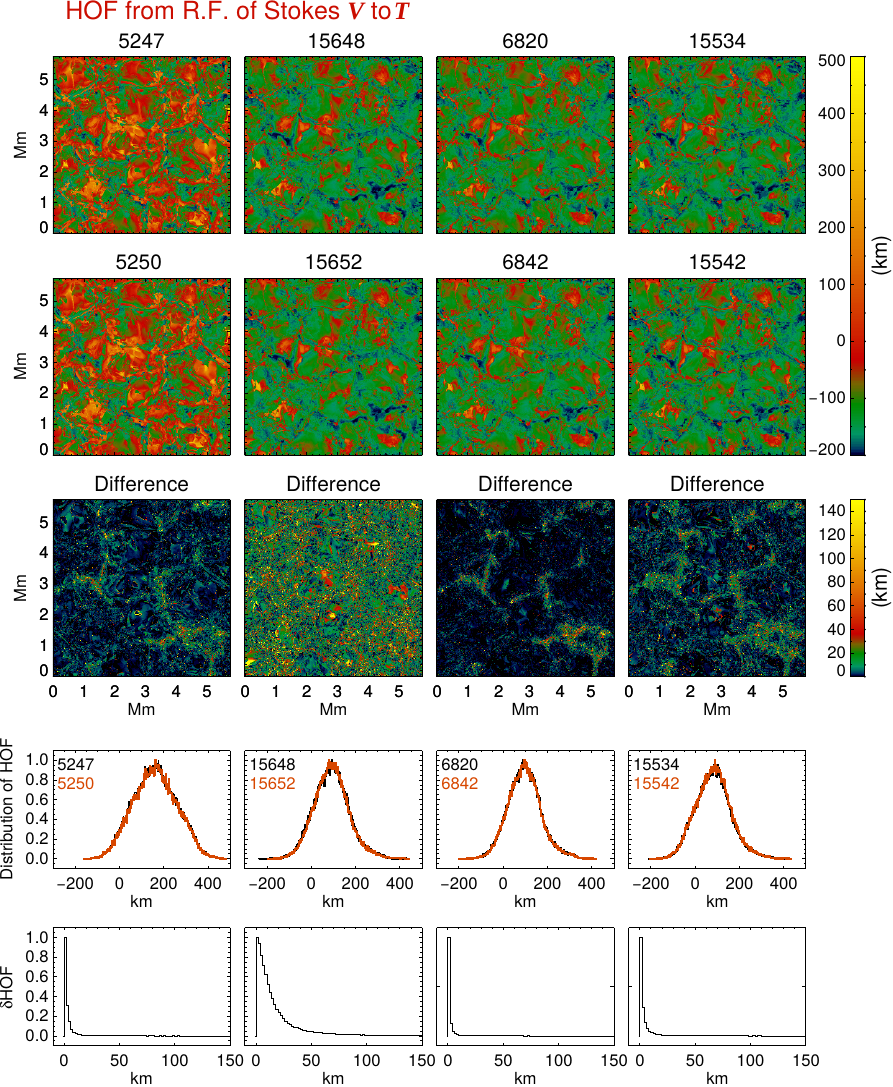}
\caption{ {Maps of height of formation for different spectral lines 
deduced from 
the centroid of the response function of Stokes $V$ profiles to temperature at the 
wavelength corresponding to the largest peak in $V$. Different rows represent the same 
quantities as in Figure~\ref{rfs}.}}
\label{rfsv}
\end{center}
\end{figure}

 {From the Stokes $V$ profiles, since the RFs are considered at 
the wavelengths of the prominent peak which is away from the line 
center, the HOFs are slightly lower in the atmosphere, especially for 
the 5250\,\AA{} pair (Figure~\ref{rfsv}). For the $1.56\, \mu$m 
pair, the distribution of the HOFs from Stokes $V$ profiles in 
Figure~\ref{rfsv} nearly overlap, unlike from the Stokes $I$ profiles 
(Figure~\ref{rfs}). Other than these difference, the overall 
distribution of the HOFs and the $\delta$HOF from $T, \varv_{\rm LOS}, 
B$ and $\gamma$ RFs are similar to the case of RFs for Stokes $I$, 
although in general the difference in the HOFs is now smaller, 
implying that the MLR should work better than suggested by 
Figures~\ref{rfs} and \ref{rfs1} alone. }

\subsubsection{New pairs}
\label{new_pairs}
Lines in the newly identified 6842\,\AA{} pair sample the same heights 
over most of the atmosphere {, evident from the maps of HOFs 
deduced from the RFs of both Stokes $I$ and $V$ profiles to 
different atmospheric properties}. Like the other pairs, this changes 
in the strong magnetic regions, with differences in HOF $>50$\,km at 
some pixels, unavoidably caused by the different Land\'e factors of 
the two lines. However, unlike the other line pairs,  the $\delta$HOF 
for the new pair always peaks at zero, with a narrow spread  {in 
all the different cases shown in Figures~\ref{rfs} -- \ref{rfsv1}}. 
This makes the line pair most suitable for the MLR method, of all the 
considered pairs, at least in this respect. Also, the pair is formed 
around 100\,km above log$(\tau_{\rm 5000})=0$ and samples deep 
photospheric layers similar to the old 1.56\,$\mu$m pair. Thus we now 
have a line pair in the visible, which can be used complementarily 
with the IR pair to probe the deep photospheric layers. In addition, 
the spread in the HOF of the 6842\,\AA{} pair is small and their 
individual RFs are quite narrow implying that they see a narrow range 
of atmospheric layers due to their higher excitation potentials, 
unlike the 5250\,\AA{} pair. This makes them less sensitive to 
magnetic field gradients, which is an advantage for the line ratio, 
but a disadvantage for their use in height-dependent inversions.
Similar is the case with the new 1.55\,$\mu$m pair. The two lines are formed at the same 
height, deep in the photosphere, as seen in the maps of HOFs from 
the RFs of both Stokes $I$ and $V$ profiles (Figures~\ref{rfs} -- 
\ref{rfsv1}). Particularly in the granules, their $\delta$HOF is 
close to zero. Due to the increased Zeeman sensitivity of the IR 
lines, this pair is well suited for the measurement of weak granular 
fields, as will be discussed in Section~\ref{mhd-mlr-section}. Also, 
notice the similarity in HOF of the new IR pair and the 15648\,\AA{} 
(g$_{\rm eff}$=3) line from the old IR pair. If two wavelength ranges 
can be covered simultaneously, then the two IR pairs, with their very 
different magnetic sensitivities, can be used together.

\section{Comparison with the 3D MHD simulations}
\label{mhd-mlr-section}
\begin{figure}
 \begin{center}
 
\includegraphics[width=0.48\textwidth]{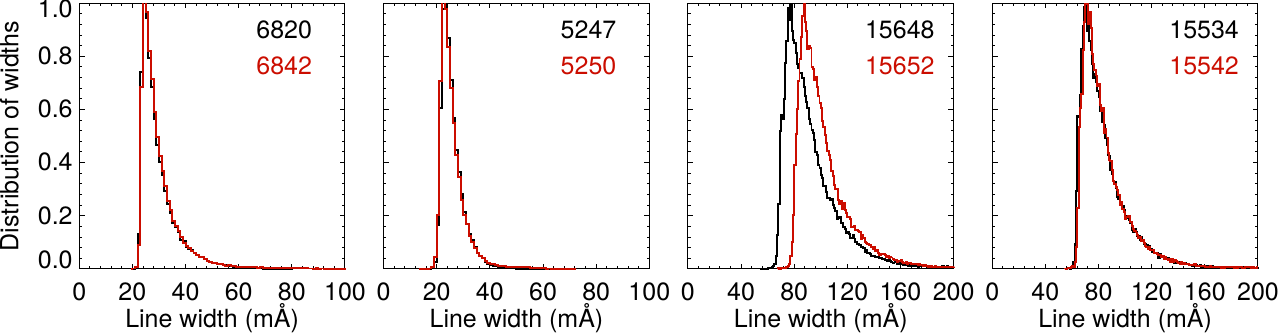}
 \caption{{Distribution of line widths in the four line pairs across the cube.}}
 \label{line_width}
 \end{center}
 \end{figure}

We define the MLR of a line pair as the ratio of Stokes $V$ amplitude 
from the magnetically weaker line (smaller g$_{\rm eff}$) to the 
stronger line (higher g$_{\rm eff}$), similar to 
\cite{2007ApJ...659.1726K}. To extract $B$ from this ratio, we need a 
calibration curve. Though neither micro- nor macro-turbulent 
velocities ($\varv_{\rm mic,mac}$) are used in the computation of the 
Stokes profiles, they are still broadened by the often strong vertical 
gradients in $\varv_{\rm LOS}$. Hence we must account for the widths 
of the spectral lines in the construction of the calibration curves. 
In Figure~\ref{line_width}, we show the distribution of the line 
widths, defined in this case as the full width at half maximum (FWHM), 
for lines in the four pairs. Except for the old $1.56\,\mu$m pair, the 
lines in each pair have practically the same line widths. The 
difference in line widths is a product of the difference in HOF 
between the lines in the $1.56\,\mu$m pair.

\begin{figure}
\begin{center}
\includegraphics[width=0.48\textwidth]{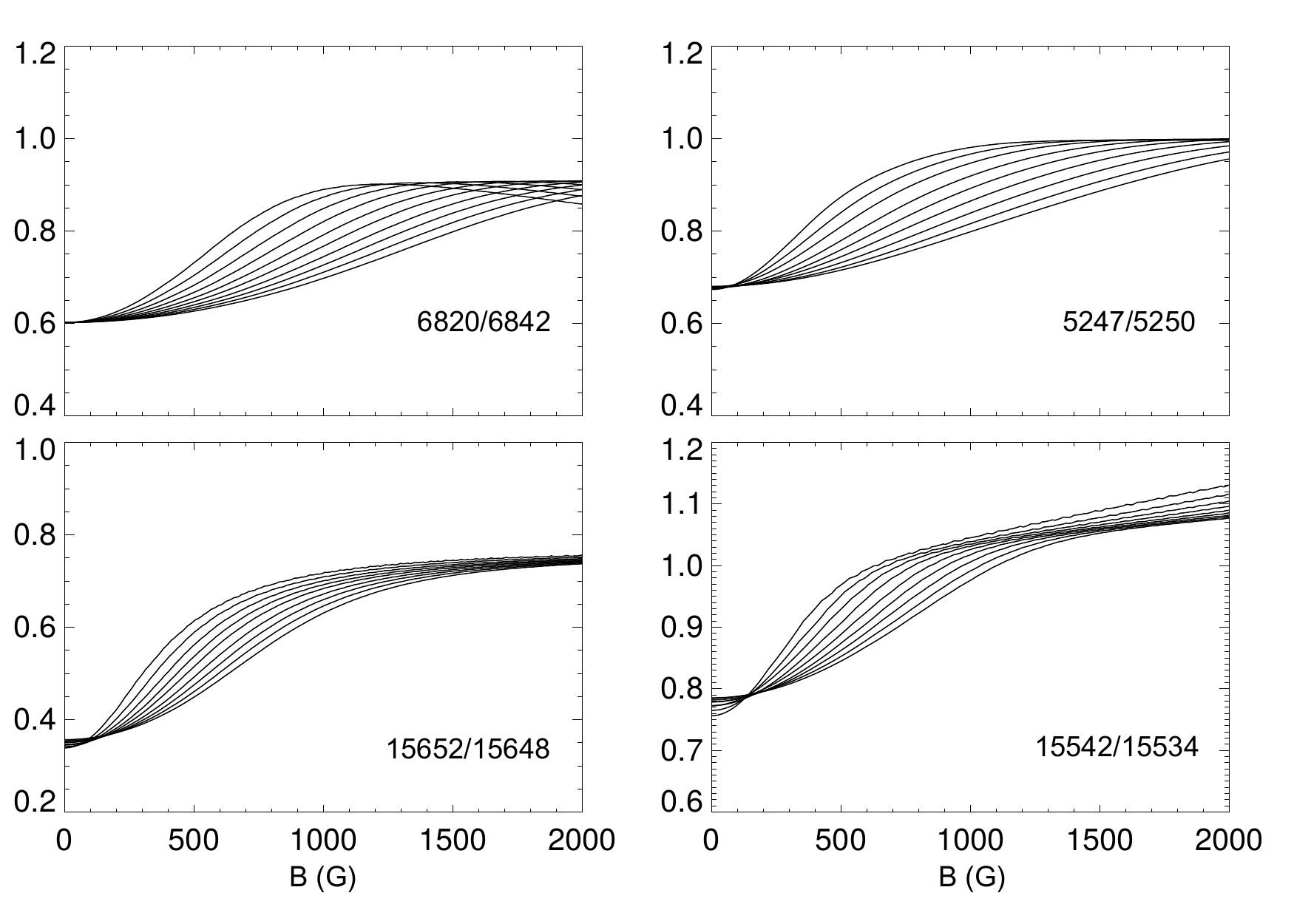}
\caption{{Ten calibration curves for each line pair, one for each of the ten bins 
into which the line widths are divided. The curves are computed for the HSRA model by 
varying height-independent micro-turbulent velocity from 0.0 km/s to 3.5 km/s.}}
\label{calib}
\end{center}
\end{figure}

Ideally, before applying the MLR, one must construct calibration 
curves at every pixel  {by fitting the intensity profiles using 
both micro and macro-turbulence}, for the four line pairs. This 
increases the number of calibration curves and they are not unique,
as different combinations of micro- and macro-turbulence are 
possible.  {In order to simplify this, we first set $\varv_{\rm 
mac}=0$ and match the line
widths using $\varv_{\rm mic}$. We then} divide the range of line 
widths into ten bins, of size 3\,m\AA{} for the visible pairs and 
10\,m\AA{} for the IR pairs. We vary a height-independent $\varv_{\rm 
mic}$ from 0.0 to 3.5\,km/s to get the required line width and 
construct a calibration curve for each width bin and each line pair. 
Figure~\ref{calib} shows the resulting calibration curves for each 
line pair. The curves are computed using the HSRA 
\citep{1971SoPh...18..347G} model atmosphere.  {Fitting both 
line width and depth by varying $\varv_{\rm mac}$ and $\varv_{\rm 
mic}$ will increase the number of calibration curves. Setting 
$\varv_{\rm mac}=0$ is a choice made to minimize the number of 
calibration curves. Despite this simplification, we recover the 
magnetic field strengths in the MHD cube relatively well, as 
discussed below.}

 {The calibration curves for the 6842\,\AA{} pair, in 
Figure~\ref{calib}, starts to decrease below the saturation level once 
the field strength exceeds a certain threshold value ($B_{\rm th}$). 
The greater the turbulent velocity or the wider the spectral line, the
higher is the value of $B_{\rm th}$. In the absence of any turbulent 
velocity (first calibration curve for the 6842\,\AA{} pair), $B_{\rm 
th} \approx 1200$\,G. Similarly, the calibration curves for the new 
$1.55\,\mu$m pair continues to increase beyond unity when the field 
strength exceeds $B_{\rm th}$ for that pair. This is because of the 
anomalous Zeeman splitting  of the spectral lines. We discuss 
this in greater detail in the appendix.}

The ambiguities involved in the comparison of $B$ from MLR with the 3D MHD cube are 
more severe than those involved while comparing the inversion results with the MHD 
simulations. The later case has been discussed in detail in \cite{2014A&A...572A..54B}. 
For the MLR, a similar comparison is made by \cite{2007ApJ...659.1726K}. In this paper, 
the authors compare the results from MLR with $B$ at log($\tau_{5000})$=-1 layer in the 
MHD cube. For only a slice of the cube, they also discuss the comparison with the fields 
weighted by the response function of Stokes $I$ to $T$. As the lines 
sample different depths across the cube  {(Figures~\ref{rfs} -- 
\ref{rfsv1})}, comparing the results of MLR with the fields at 
constant $\tau$ will not properly indicate the reliability of the line 
pair. Hence, we discuss below a different way of comparing $B$ from 
MLR and the MHD cube. 

{Traditionally, the MLR is computed by either taking the ratio of the blue 
Stokes $V$ peak, as done in, e.g. \citet[][]{2007ApJ...659.1726K, 
2010A&A...517A..37S, 2013A&A...556A.113S}, or by taking the sum or the average of the 
blue and the red lobes \citep[e.g.,][]{1985SoPh...95...99S, 1987A&A...188..183S}. The 
rationale behind the former is that the blue peak is less affected by magnetic and 
velocity gradients, and that they have larger amplitudes \citep{2010A&A...517A..37S}. 
Taking the sum or the average of the blue and the red lobes improves the signal to noise 
ratio. However, while comparing with the magnetic field strength in the MHD cube, we find 
that the ratio of the most prominent peak (the lobe with higher amplitude) in the Stokes 
$V$ profile performs better than the other two ratios. A similar approach has also been 
followed by \cite{2016A&A...596A...6L}.}

\begin{figure*}
\begin{center}
\includegraphics[width=0.8\textwidth]{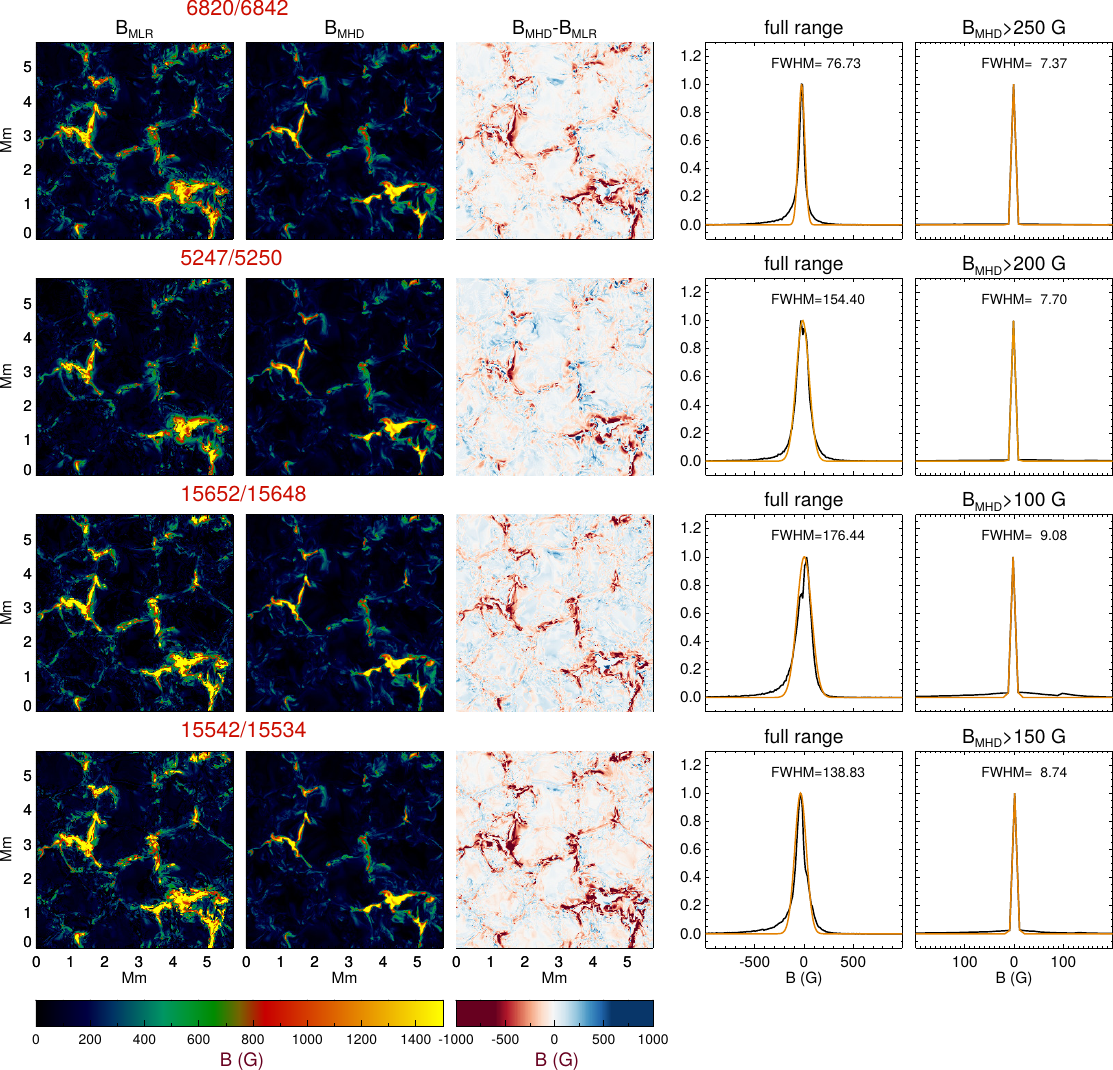}
\caption{Comparison between the magnetic field strengths from the MLR ($B_{\rm MLR}$) and 
the MHD cube ($B_{\rm MHD}$). The fields in the MHD cube are computed using 
Equations~(\ref{bmhd1}) and (\ref{rf-2}). The third column is the difference, $B_{\rm MHD} 
- B_{\rm MLR}$. The fourth column is the histogram of the difference (black curve) which 
is fit with a Gaussian (brown curve) and the width of the Gaussian is indicated in each 
case. Fifth column is same as the fourth but only over the field range where the MLR is 
most effective.}
\label{mhd-mlr}
\end{center}
\end{figure*}

If $B$ from MLR is $B_{\rm MLR}(x,y)$ and $B$ in the MHD cube is $B_{\rm 
MHD}(x,y)$, then  
\begin{equation}
B_{\rm MHD}(x,y) = \frac{\int{{\rm{RF\_tot}}^{V}_{B}(x,y,\tau)}\  B(x,y,\tau)\  
d\tau}{\int{{\rm{RF\_tot}}^{V}_{B}(x,y,\tau)}\  d\tau},
\label{bmhd1}
\end{equation}
where ${\rm RF\_tot}^{V}_{B}(x,y,\tau)$ is the total RF for the two lines defined as
% \begin{eqnarray}
% {\rm RF\_tot}^{V}_{B}(x,y,\tau) = \int{\rm RF}^{V}_{B}(x,y,\tau,\lambda_1)
% d\lambda_1 \nonumber \\ + \int {\rm RF}^{V}_{B}(x,y,\tau,\lambda_2) d\lambda_2,
% \label{rf-1}
% \end{eqnarray}
% \begin{eqnarray}
% {\rm RF\_tot}^{V}_{B}(x,y,\tau) = \int{\rm RF}^{V}_{B}(x,y,\tau,\lambda_1)
% d\lambda_1 + \int {\rm RF}^{V}_{B}(x,y,\tau,\lambda_2) d\lambda_2,
% \label{rf-1}
% \end{eqnarray}

\begin{equation}
{\rm RF\_tot}^{V}_{B}(x,y,\tau) = {{\rm RF}^{V}_{B}(x,y,\tau,\lambda_{p1}) + {\rm 
RF}^{V}_{B}(x,y,\tau,\lambda_{p2})}.
\label{rf-2}
\end{equation}

In Equation~(\ref{rf-2}), ${\rm{RF}}^{V}_{B}(x,y,\tau,\lambda_{p1,p2})$ are the RFs at 
wavelengths, $\lambda_{p1,p2}$, corresponding to the peak value of the Stokes $V$ 
profile from the MHD cube. It is this peak value which is then used to compute the MLR. 

In Figure~\ref{mhd-mlr}, we show the comparison between $B_{\rm MLR}$ (first 
column) and $B_{\rm MHD}$ (second column) computed using Equations~(\ref{bmhd1}) and 
(\ref{rf-2}). The third column is the difference, $B_{\rm MHD}-B_{\rm 
MLR}$. The last two columns depict the histograms of the differences over the full range 
of $B_{\rm MHD}$ (fourth column) and over the range where the MLR method is most 
effective (fifth column). The latter is plotted starting from the field strengths for 
which  the more Zeeman sensitive line of each pair enters the non-linear Zeeman regime, or 
in other words, from where the calibration curves start to have a steep gradient. We then 
apply a Gaussian fit to the histogram and the FWHM of the Gaussian curve is indicated for 
each pair.

The fields in the MHD cube are well reproduced by all the four line pairs.  The 
differences between the $B_{\rm MLR}$ and $B_{\rm MHD}$ seen in the third column resemble 
the $\delta$HOF images in Figures~\ref{rfs} and \ref{rfs1}. When the 
full range of field strengths are considered, the scatter is the smallest from the 
6842\,\AA{} pair and largest from the old $1.56\,\mu$m pair. When the reliability of the 
pairs are tested over the field strength range where they are most efficient, all the 
line pairs perform equally well and the scatter is very small. 

The difference image in the third column which covers the full range of field strengths, 
has contributions from three factors: First is from those pixels where the fields are 
weak and the lines are still in the weak field regime, i.e., the Zeeman splitting 
is much smaller than the Doppler width. Hence in the fifth column, we show the histogram 
of the difference by excluding these weak fields. The 6842\,\AA{} pair and the 5250\,\AA{} 
pair are in this regime up to $\approx 250\,$G. This is seen from the calibration curves 
in Figure~\ref{calib}. Here, the Stokes $V$ ratio is equal to the ratio of g$_{\rm eff}$ 
of the two lines. However, the IR pairs are in the weak field regime 
for field only up 
to $\approx 100-150$\,G. Hence, they can measure weak granular fields better than the 
visible pairs. Among the two IR pairs, the new $1.55\,\mu$ pair performs even better 
in the granules because of the same HOF of the two lines. This is indicated by the white 
patches seen in the difference image at the granules. 

The second factor contributing to the difference is the increase in $\delta$HOF in the 
regions surrounding strong magnetic field concentrations, due to the canopies 
(Section~\ref{new_pairs}). From Figures~\ref{rfs} -- \ref{rfsv1}, this increase is seen 
in all the four line pairs. In these regions, the $B_{\rm MLR} > B_{\rm MHD}$ and such 
locations are seen as brown patches surrounding strong field regions in the difference 
images of Figure~\ref{mhd-mlr}, noted also by \citet{2007ApJ...659.1726K}. 
{Contributions from these pixels to the histogram of the difference (fourth column 
in Figure~\ref{mhd-mlr}) appear in the left wing of the Gaussian which extends up to 
1000\,G. These pixels do not contribute to the histogram in the fifth column because they 
are constructed by imposing criteria on $B_{\rm MHD}$. The $B_{\rm MHD}$ in these pixels 
are below the imposed criteria.} Differences in $B_{\rm MLR}$ and $B_{\rm MHD}$ are also 
seen along the edges of the granules, i.e along the granular-intergranular boundaries. 
Once again, this is due to the increase in $\delta$HOF caused by 
strong $T, \varv_{\rm LOS}$, and $B$ gradients, seen from 
Figures~\ref{rfs} -- \ref{rfsv1}. 

The third factor is the saturation (or near saturation) of the calibration curves for 
stronger field strengths. The calibration curves for the visible line pairs, for larger 
line widths, do not saturate even at 2000\,G (Figure~\ref{calib}). For the IR pair, 
the calibration curves saturate around 1200\,G. 

\section{MLR with degraded profiles}
\label{spdg-section}
In the previous section, we discussed the line pairs and MLR under ideal conditions 
but in reality, the observations from any instrument are affected by noise and atmospheric 
seeing. In this section we discuss the influence of these effects on the line profiles and 
the results from MLR.

\begin{figure}
\begin{center}
\includegraphics[width=0.48\textwidth]{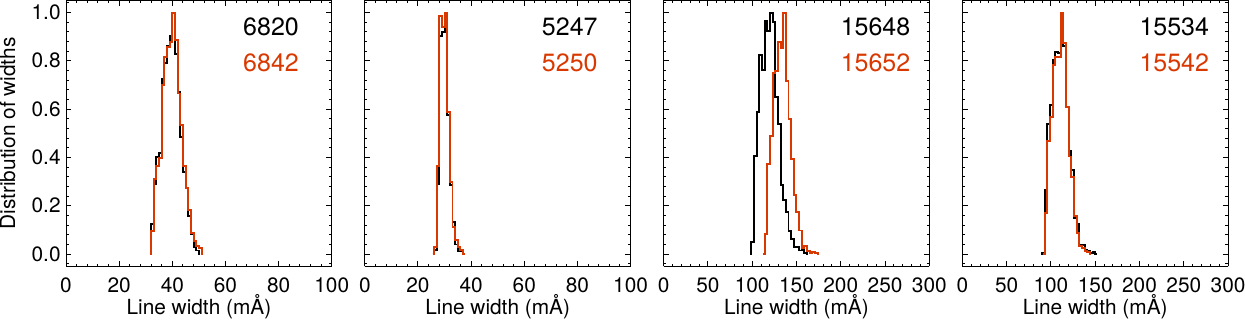}
\caption{Distribution of line widths in the four line pairs across the cube 
after the Stokes profiles are spectrally and spatially degraded.}
\label{line_width_spdg}
\end{center}
\end{figure}

\begin{figure}
\begin{center}
\includegraphics[width=0.48\textwidth]{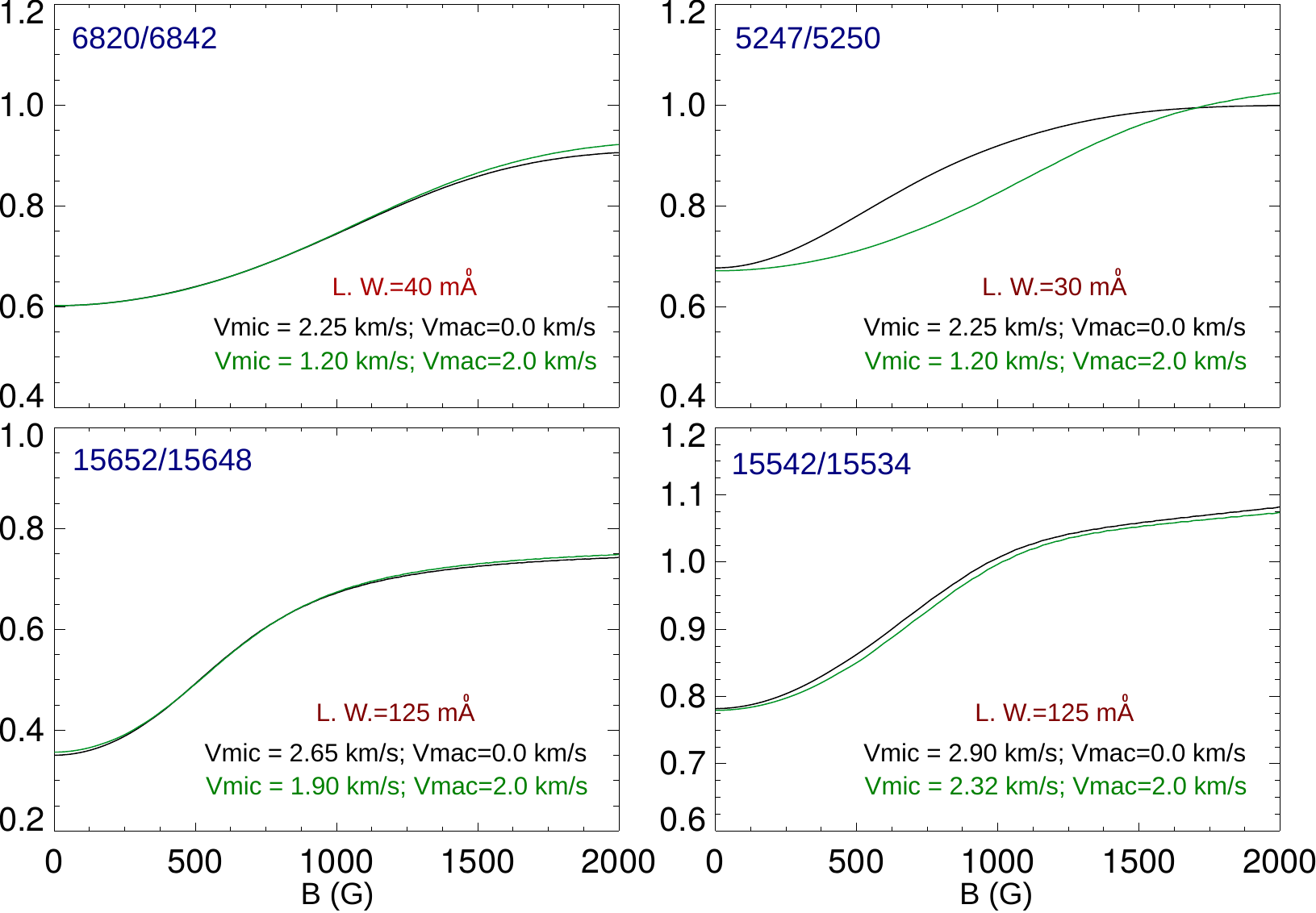}
\caption{{Calibration curves for the four line pairs shown for sample line 
widths indicated by L.W.  The black and the green curves are with and without a 
macro-turbulent velocity of 2\,km/s. In the green curves, the micro-turbulent velocity is 
reduced, to get the same line width.}}
\label{calib_vmac}
\end{center}
\end{figure}

\begin{figure*}
\centering
\includegraphics[width=0.8\textwidth]{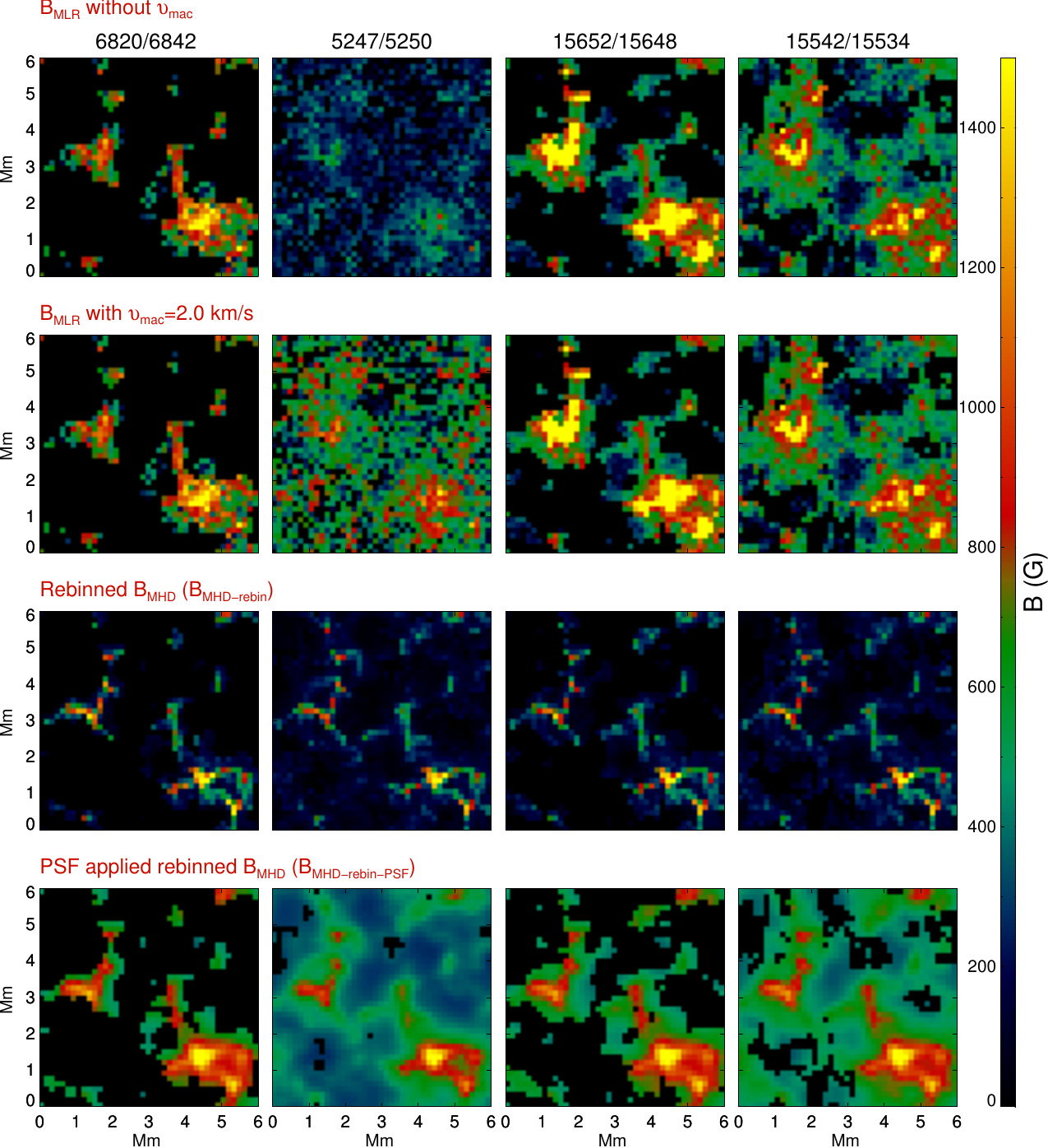} 
\caption{$B_{\rm MLR}$ determined from spatially and spectrally 
degraded Stokes $V$ profiles affected by noise, using calibration 
curves without macro-turbulent velocity (top row), and with a 
macro-turbulent velocity of 2.0\,km/s (second row). 
 {This is compared with the magnetic field strength $B_{\rm 
MHD}$ re-binned according to Equation~\ref{spdg-eqn1} ($B_{\rm 
MHD-rebin}$, third row). We show an empirical approximation of the 
result of MLR, based on $B_{\rm MHD-rebin}$ and including the 
influence of the PSF in the fourth row Note that all profiles have 
been broadened with a micro-turbulence to match the widths of the 
profiles emerging from the MHD snapshot.}}
\label{spdg-mlr}
\end{figure*}

To simulate the solar observations, we apply both spatial and spectral 
degradation to the Stokes profiles from the MHD cube and then estimate 
the $B_{\rm MLR}$. For this, the synthesized Stokes profiles are 
convolved with the theoretical point spread function (PSF) of the 
GREGOR telescope which includes the effects of spatial stray light. 
The profiles are then spectrally degraded by convolving them with a 
Gaussian with FWHM=30\,m\AA{} and 100\,m\AA{} respectively for the 
visible and IR line pairs. Later, they are re-binned to a detector 
pixel resolution of $0\arcsec.2$. For further details on the PSF used, 
see \citet{2016A&A...596A...6L}. 

In addition to the degradation,  we add a random noise of $\sigma=1\times10^{-3}$ in 
the units of continuum intensity of the respective pair. 
We then consider all profiles with an amplitude larger than $3\sigma$ and apply
a median filter over three wavelength pixels to smoothen the Stokes profiles. This 
threshold is applied to the magnetically weaker of the two lines in 
the pair.  After spectral 
degradation and filtering, the Stokes profiles are further broadened. Hence we must 
construct new set of calibration curves for the four pairs. Repeating the same procedure 
as before, we plot the histograms of the line widths over the whole cube, in 
Figure~\ref{line_width_spdg}. The line widths of the profiles are grouped into bins of 
3\,m\AA{} and 10\,m\AA{} for the visible and IR pairs, respectively. At first, only 
$\varv_{\rm mic}$ is varied to match the line widths while keeping $\varv_{\rm 
mac}=0$, and the calibration curves are constructed. The effects of $\varv_{\rm mac}$ will 
be discussed later in the section.

The MLR estimates $B$ within the resolution element irrespective of 
the filling factor. In other words, in a resolution element 
containing 
a mix of magnetic and nonmagnetic components or strong and weak 
magnetic components, the MLR measures $B$ mainly from the strong magnetic 
component in the element and not the spatially averaged $B$ 
\citep{1973SoPh...32...41S}. Hence, to compare with the fields in the 
MHD cube, we weight $B$ with the $V$ amplitude. By doing so, 
we give more weight to the magnetic field at locations where the 
Stokes V profile is stronger. In general these are the stronger 
magnetic fields (aligned along the line of sight), while the weaker 
(or more transverse) fields provide a proportionally smaller 
contribution to the line ratio. When such a weighted magnetic field 
strength is averaged to match the degraded pixel resolution, the 
resulting field strength has contributions mainly from the stronger 
magnetic component and resemble the field strength measured by MLR. 
Below we give an empirical relation aiming to provide a magnetic 
parameter that approximates the field strength sampled by the line 
ratio technique in the presence of finite spatial resolution. We call 
the magnetic field computed using this relation as $B_{\rm 
MHD-rebin}$. It is given by
\begin{eqnarray}
{B_{\rm MHD-rebin}(x^{\prime}_p, 
y^{\prime}_q)}\ {=}\ {\frac 
{\sum\limits_{j=aq}^{a(j+1)-1} \sum\limits_{i=ap}^{a(i+1)-1} [B_{\rm 
MHD}(x_i,y_j) V(x_i,y_j)]}{\sum\limits_{j=aq}^{a(j+1)-1} 
\sum\limits_{i=ap}^{a(i+1)-1} [V(x_i,y_j)]}},
\label{spdg-eqn1}
\end{eqnarray}
where $p=0, 1, .., m-1$; $q=0, 1, .., n-1$, and the summations 
rebin the quantities in the square brackets. For the present purposes, the 
rebinning is done over 7 pixels, i.e. $a=7$, to match a detector pixel 
resolution of $0\arcsec.2$. The dimensions of $B_{\rm MHD-rebin}(x^{\prime}, 
y^{\prime})$ is $(m,n)$. In the above 
equation, $B_{\rm MHD}(x,y)$ is computed from Equations~(\ref{bmhd1}) 
and (\ref{rf-2}) . $V(x,y)$ is the amplitude of the Stokes $V$ 
profile at pixel $(x,y)$ from the MHD cube at full resolution. When 
the fields are weak, $V \propto B$ and from Equation~\ref{spdg-eqn1}, 
$B_{\rm MHD-rebin}$ is $B_{\rm MHD}$ averaged over the resolution 
element.

Figure~\ref{spdg-mlr} shows a comparison between the $B_{\rm MLR}$ 
from the spatially and spectrally degraded profiles with $B_{\rm 
MHD-rebin}$ defined in Equation~\ref{spdg-eqn1}. In the first row the 
$B_{\rm MLR}$ is computed from the calibration curves which are 
constructed by varying only $\varv_{\rm mic}$ to match the line 
widths. The maps in the second row will be discussed later. 
The third row shows the magnetic field maps resulting from 
Equation~\ref{spdg-eqn1}. The shapes of the magnetic field structures 
from the MLR method in the first row do not resemble those in the 
third row. This is because the magnetic field structures in the first 
row, which are obtained by applying the MLR method on the PSF 
convolved Stokes $V$ profiles, are smeared out. This effect has not 
been accounted for, in Equation~\ref{spdg-eqn1}. In order to 
reproduce this effect, we apply the PSF to $B_{\rm MHD-rebin}$, to get $B_{\rm 
MHD-rebin-PSF}$. We stress that there is no clearcut physically consistent 
manner in which $B_{\rm MHD-rebin}$ can be convolved with the PSF. Our aim here 
is to empirically get a better idea of what quantity the MLR actually returns in 
a realistic atmosphere in the presence of spatial smearing. To that end we 
tried different things and compared the resulting maps with the first row of 
Figure~\ref{spdg-mlr}. We found that the best agreement (in the shape of the 
features) was obtained by
\begin{eqnarray}
B_{\rm MHD-rebin-PSF}(x^{\prime}, 
y^{\prime})&=&{\rm 
PSF}(x^\prime,y^{\prime})\ \ast\ B_{\rm 
MHD-rebin}(x^{\prime},y^{\prime}),
\label{spdg-eqn2}
\end{eqnarray}
where $B_{\rm MHD-rebin}(x^{\prime},y^{\prime})$ is defined in 
Equation~\ref{spdg-eqn1} and $\ast$ represents convolution. After applying the 
PSF, however, the magnetic field is smeared and diluted 
\citep{2016A&A...596A...6L}. Thus $B_{\rm MHD-rebin-PSF}$ is much smaller than 
$B_{\rm MHD-rebin}$ (third row of Figure~\ref{spdg-mlr}). In the presence of 
spatial smearing, though result from MLR is spatially smeared, the strength of 
the field is maintained (i.e. still the intrinsic field strength is reached at 
the centers of magnetic features) and thus the $B_{\rm MHD-rebin-PSF}$ from 
Equation~\ref{spdg-eqn2} is smaller also than the MLR results shown in the top 
row of Figure~\ref{spdg-mlr}. Hence we normalize $B_{\rm MHD-rebin-PSF}$, such 
that its maximum field strength matches with the maximum of $B_{\rm MHD-rebin}$. 
In the fourth row we show maps of the normalized $B_{\rm MHD-rebin-PSF}$ and the 
pixels where the degraded Stokes $V$ is smaller than 3$\sigma$ are filtered 
out. Now the field structures in the first row resemble those in the fourth row.

The $B_{\rm MHD}$ in Equation~\ref{spdg-eqn1} is obtained after weighting the 
original field in the MHD cube with the response function and integrating over 
tau, from Equations~\ref{bmhd1} and \ref{rf-2}. Therefore, the original 
intrinsic field strength in the MHD cube is maintained. With 
Equations~\ref{spdg-eqn1} and \ref{spdg-eqn2}, we are trying to empirically 
represent the quantity that MLR method provides in a realistic atmosphere and 
for realistic instrumental degradation. This is not straightforward and has not 
been reported in the literature. By comparing the maps in the first and the 
fourth rows in Figure~\ref{spdg-mlr}, we see that this empirical representation 
provides a reasonably close match with $B_{\rm MLR}$.

Due to smaller $V$ amplitudes in the 6842\,\AA{} line pair and the 
$1.56\,\mu$m line pair, about 30--45\% of the profiles are above the 
$3\sigma$ threshold. As the lines in the 5250\,\AA{} pair and the 
$1.55\,\mu$m pair are stronger, more than 80\% of the profiles remain 
above the $3\sigma$ level. The 6842\,\AA{} pair and the two IR pairs 
clearly show the presence of kG fields in the cube. But they are 
spread over larger areas because of the convolution with the PSF. The 
green patches surrounding the strong field yellow patches are due to 
redistribution of the photons caused by the PSF. This is also 
discussed in detail by \citet{2016A&A...596A...6L}.

The 5250\,\AA{} line pair, however, does not measure kG fields in the 
cube (first row in Figure~\ref{spdg-mlr}). From this line pair, kG 
fields were not recovered also by \citet{2007ApJ...659.1726K} in an 
MHD cube and by \citet{2008ApJ...674..596S} in  solar network 
observations. In the former paper, the authors explained this to be 
due to the larger formation heights of the lines in the 5250\,\AA{} 
pair and that they sample weaker magnetic fields in the MHD cube. As 
kG fields could not be recovered in the network observations by 
\citet{2008ApJ...674..596S}, they concluded this line pair to be 
unreliable and that it is no better than the 6300\,\AA{} pair in which 
the two lines are formed at very different heights in the atmosphere. 
This is surprising because, the presence of kG fields in the solar 
network regions was discovered by applying MLR to the 5250\,\AA{} line 
pair by \citet{1973SoPh...32...41S}.

{To investigate this, we included a constant height-independent 
$\varv_{\rm mac}$ of 2\,km/s in addition to $v_{\rm mic}$ and 
recomputed the calibration curves. The $\varv_{\rm mic}$ was varied to 
get the required line widths. Examples comparing the calibration 
curves with and without $\varv_{\rm mac}$ for fixed line widths is 
shown in Figure~\ref{calib_vmac}. The 5250\,\AA{} line pair is the 
most affected by the addition of $\varv_{\rm mac}$. This pair is 
highly sensitive to both $\varv_{\rm mic}$ and $\varv_{\rm mac}$, as 
pointed out in \citet{1987A&A...188..183S,2007ApJ...659.1726K}. The 
magnetic field strengths recovered from the new calibration curves are 
shown in the second row of Figure~\ref{spdg-mlr}. The 5250\,\AA{} pair 
now shows the presence of kG fields in the cube. However, the 
magnetic field map from this line pair does not match well with those 
in the fourth row. This could be because of the approximations in the 
construction of the calibration curves. If the curves are constructed at every 
pixel by fitting the full spectral line then the 5250\,\AA{} pair may provide a 
better comparison with the magnetic field maps in the fourth row. The results 
from the other three line pairs are not much affected by the addition of 
$\varv_{\rm mac}$, as also seen from Figure~\ref{calib_vmac}. What we have 
presented is only a simplified approach, so that, if the 5250\,\AA{} line pair 
is to be used for MLR, both $\varv_{\rm mic}$ and $\varv_{\rm mac}$ should 
be varied to match the line width and depth at every pixel in the cube. In any 
case, the 5250 line pair is less robust than the others.}

\section{Conclusions}
\label{conclusions}
The magnetic line ratio (MLR) method has been widely used to measure magnetic field 
strengths on the Sun. Until recently, three line pairs (5250\,\AA{}, 6300\,\AA{} and 
$1.56\,\mu$m pairs) were used for this method, only two of which (5250\,\AA{} and 
$1.56\,\mu$m pairs) give reliable results. In this paper, we have identified two new line 
pairs, the 6842\,\AA{} pair in the visible and the $1.55\,\mu$m pair in the IR. Lines in 
the 6842\,\AA{} pair are separated by 22\,\AA{} and those in the new $1.55\,\mu$m pair by 
8\,\AA{}. Lines in each of these pairs are formed at roughly the same height in the 
atmosphere. The new pairs have one line with high g$_{\rm eff}$ and with large difference 
in g$_{\rm eff}$ between the lines, making them well suited for MLR.  We have presented a 
detailed comparison of the new and the old line pairs. 

The Stokes profiles are synthesized in a three dimensional MHD cube having a 
field 
strength $B_{\rm MHD}$ (which differs from one pixel to the next). The MLR method is 
applied to the synthesized profiles to recover the field strengths, called $B_{\rm MLR}$. 
The $B_{\rm MLR}$ compares well when the $B_{\rm MHD}$ is weighted with the 
Stokes $V$ response function and then integrated over the optical depth grid. 
All the four line pairs reproduce $B_{\rm MHD}$, but the scatter in histogram of 
the difference between $B_{\rm MHD}$ and $B_{\rm MLR}$ is smaller for the new 
visible and IR pairs. The two lines in the new IR pair are  stronger than the 
lines in the old $1.56\,\mu$m pair. Though the lines in new IR pair have Stokes 
$V$ signals that are typically smaller than the 15648\,\AA{} line (g$_{\rm 
eff}=3$ line), they are much stronger than those of the g$_{\rm eff}$= 1.53 line 
at 15652\,\AA{} line, used together with $\lambda$ 15648\,\AA. Thus, in the 
presence of noise, the Stokes profiles of both lines in the new $1.55\,\mu$m 
pair will remain above noise more often than the $1.56\,\mu$m pair, making them 
favourable also for the inversions.

We have further tested the line pairs by applying spatial and spectral 
degradation, and by adding random noise ($\sigma = 1 \times 10^{-3} 
I_{c}$) to the Stokes profiles. We find that the new 6842\,\AA{} pair 
and the old $1.56\,\mu$m pair are most affected by noise. However, 
more than 80\% of the Stokes $V$ profiles from the new IR pair, remain 
above the $3\sigma$ cutoff. 

{Using the 5250\,\AA{} line pair, \citet{2007ApJ...659.1726K} and 
\citet{2008ApJ...674..596S} could not recover kG fields from the profiles synthesized in 
a 3D MHD cube and in the solar network observations, respectively. While 
\citet{2007ApJ...659.1726K} attributed this to the larger formation heights of the lines 
in the 5250\,\AA{} pair, \citet{2008ApJ...674..596S} concluded this line pair to be 
unreliable. We find that the 5250\,\AA{} pair is more sensitive to the nature of the 
velocity field, e.g. the exact mixture of micro and macro-turbulent velocities, than the 
other line pairs. Also, since the lines in this pair are strong and temperature sensitive, 
it is necessary to match the full line shape (line width and line depth) in the 
construction of calibration curves. From the calibration curves with the right 
combination of micro and macro-turbulent velocities, it is possible to measure 
kG fields from the 5250\,\AA{} pair. For the other three line pairs (6842\,\AA, 
old $1.56\,\mu$m and new $1.55\,\mu$m pairs), calibration curves constructed by 
matching the line widths is sufficient for measuring reliable magnetic field 
strengths.}

{The interpretation of the MLR has in the past been generally given in terms of 
an idealized 2-component atmosphere, a field free and a homogeneous magnetic 
component \citep{1973SoPh...32...41S}. In this representation the field strength 
returned by the MLR is an approximation of the intrinsic field strength in the 
magnetic component. What happens in a more realistic, complex atmosphere with a 
distribution of field strengths and the influence of a PSF? Here it turns out 
that the MLR still gives an approximation of the intrinsic field strength at the 
average formation height of the Stokes $V$ lobes, but weighted by the amplitude 
of the Stokes $V$ profile (regions with small Stokes $V$ provide a smaller 
contribution). Also the influence of spatial smearing turns out to be complex. 
Ours is the first attempt to empirically determine what exactly the MLR returns 
in a realistic atmosphere. It can likely be improved. }

Sophisticated inversion codes are currently the preferred choice for magnetic 
field measurements. We expect the new line pairs to be attractive pairs also for the 
application of inversion codes. In addition, it may be possible to combine the MLR with 
the inversions. One way would be to use the magnetic field strength measured from MLR as 
an initial guess in the inversions. Another is to employ the MLR as an additional 
constraint on the inversion. This will be investigated in a forthcoming paper. 

\begin{acknowledgements}
 {We thank the referee for useful comments and suggestions 
which helped in improving the paper.} The authors are grateful to 
L.~P.~Chitta, A.~Lagg, I.~Mili\'{c} and 
M.~van Noort for helpful discussions. Thanks to T. Riethm\"{u}ller for 
kindly providing the MURaM MHD cube. HNS acknowledges the financial 
support from the Alexander von Humboldt Foundation.  {This 
project has received funding from the European Research Council (ERC) 
under the European Union’s Horizon 2020 research and innovation 
programme (grant agreement No. 695075) and has been supported by the 
BK21 plus program through the National Research Foundation (NRF) 
funded by the Ministry of Education of Korea. This research has made 
use of NASA's Astrophysics Data System.}
\end{acknowledgements}

%\bibliography{line-ratio}

\begin{appendix}

\section{Anomalous Zeeman splitting and MLR}
 \begin{figure*}[b]
\begin{center}
\includegraphics[width=0.8\textwidth]{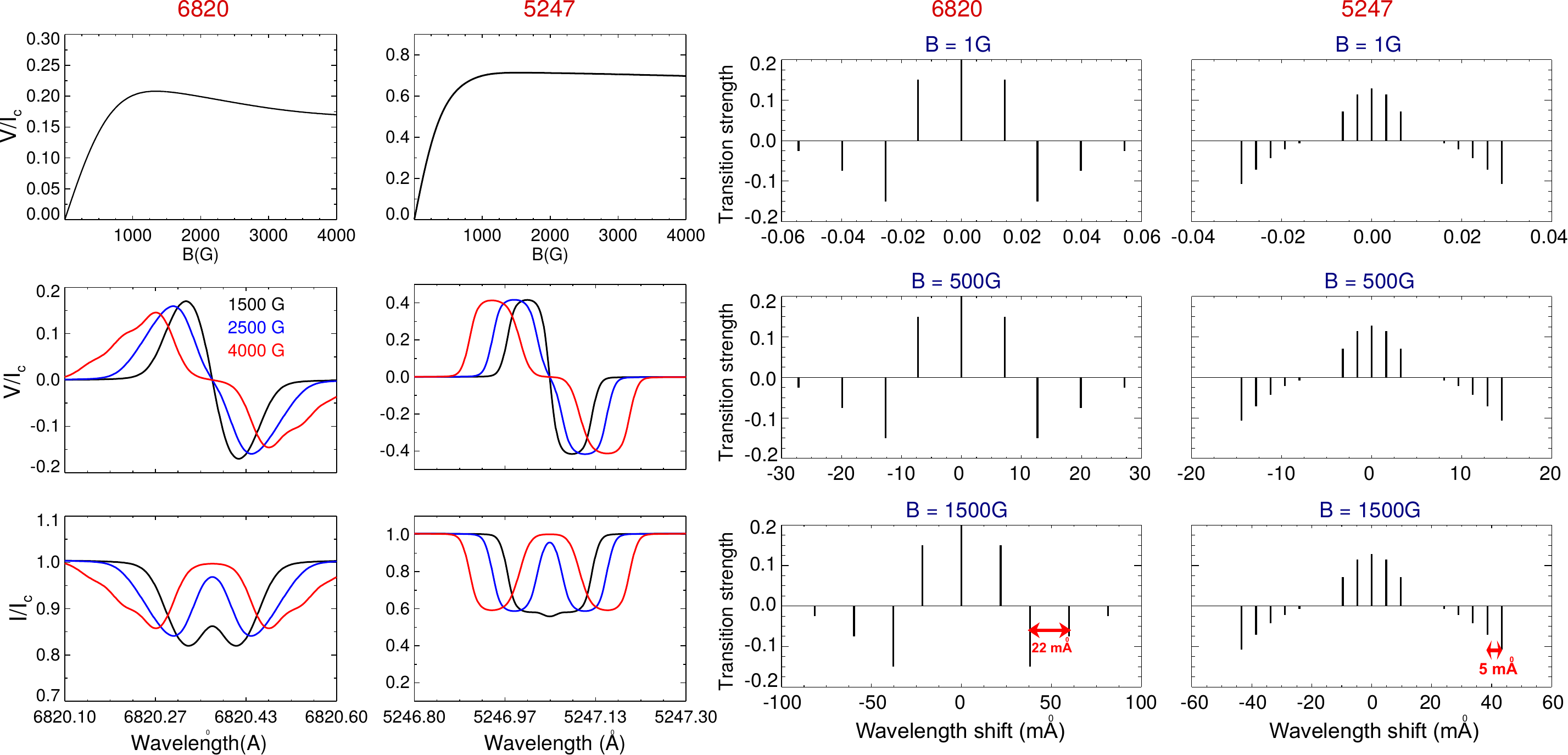}
\caption{ {Variation in Stokes $V$ amplitude for the 
6820\,\AA{} and 5247\,\AA{} lines as a function of magnetic field strength 
($B$). Full line profiles are shown for 1.5, 2.5 and  4\,kG. A comparison 
between the Zeeman splitting pattern for the two lines for $B$= 1, 500 and 
1500\,G is shown in the last two columns. Note the change in the wavelength 
scale of these plots with $B$.}}
\label{zeeman}
\end{center}
\end{figure*}

In the presence of strong magnetic fields, the Zeeman saturation suppresses the 
amplitude of a Stokes $V$ profile and broadens its lobes. In normal Zeeman 
triplets, as $B$ increases, the amplitude of Stokes $V$ increases until it 
saturates. For field strengths beyond that, it remains constant. When the 
Zeeman splitting is anomalous, Stokes $V$ continues to change with $B$ 
\citep{1993SSRv...63....1S}. In the new line pairs, the 6820\,\AA{} and 
15534\,\AA{} lines undergo anomalous Zeeman splitting. For stronger fields, 
their Stokes $V$ amplitudes decrease when $B$ exceeds a certain threshold value, 
$B_{\rm th}$.

 {The 5247\,\AA{} line is also not a normal Zeeman triplet, but 
its Stokes $V$ begins to decrease significantly only when $B$ 
exceeds 5\,kG. A comparison between the 6820\,\AA{} and the 
5247\,\AA{} lines is shown in Figure~\ref{zeeman}. The behaviour of 
the Stokes $V$ amplitude is governed by the splitting of the 
individual transitions forming a given $\sigma$-component. For 
$B=1.5$\,kG, the separation between the various transitions in a 
$\sigma$-component ($\Delta\lambda_B$) in the 6820\,\AA{} line is 
as high as 22\,m\AA, whereas in the 5247\,\AA{} line it is only 
5\,m\AA{} (indicated with red arrows in Figure~\ref{zeeman}). As $B$ 
increases, $\Delta\lambda_B$ becomes comparable to the line widths of 
the 
individual transitions, resulting in broadening of the 
$\sigma$-component and a corresponding decrease of the $V$ amplitude. 
This is clear when $B$ is increased to 4\, kG, we see peaks of the 
line profiles from each transition in the 6820\,\AA{} line (first 
column in Figure~\ref{zeeman}) but not in the 5247\,\AA{} line (second 
column in Figure~\ref{zeeman}).}

 {The $\Delta\lambda_B$ for a Zeeman component 
($\pi$ or $\sigma$) is proportional to ($m_l$\,g$_l\, -\, m_u$\,g$_u$) 
where $m_{l,u}$ are the magnetic quantum numbers of the lower and 
upper levels of the transition, respectively. The Land\'{e} g-factors 
of the upper and lower atomic levels are denoted as g$_{u}$  and 
g$_{l}$, respectively. For the $\sigma$ components, $\delta m = (m_u\, 
-\, m_l)$ is $\pm 1$ and hence $\Delta\lambda_B \propto m_u 
($g$_l\,-\,$g$_u)\ \pm\ $g$_l$ \citep{2007insp.book.....D}. For 
the 6820\,\AA{} line, $\Delta\lambda_B$ is much larger with 
$\delta$g =|g$_l\,-\,$g$_u$|= 0.67 and g$_l$=2.5 compared to the 
5247\,\AA{} line, with $\delta$g = 0.25 and g$_l$=0.5. The 
$\Delta\lambda_B$ is large also for the 15534\,\AA{} line 
with $\delta$g = 0.33 and g$_l$=1.5. If 
g$_l$ = 0 or g$_u$ = 0 or $\delta$g = 0 then it is a normal Zeeman 
triplet and there is no change in $V$ amplitude after Zeeman 
saturation. The calibration curves for MLR for line pairs having 
at least one line which undergoes anomalous Zeeman splitting do not 
saturate, that is, reach a constant value for stronger fields. 
Depending on whether the line with anomalous splitting is the 
magnetically weaker or the stronger in the pair, the calibration 
curve, when computed as the ratio of magnetically weaker to the 
stronger line, either decreases (6842\,\AA{} pair) or increases 
(1.55\,$\mu$m pair) with $B$ as seen from Figure~\ref{calib}.}

 {The profiles in Figure~\ref{zeeman} are computed without 
$\varv_{\rm mac}$ and $\varv_{\rm mic}$. Velocity broadening increases 
the value of $B_{\rm th}$ at which Stokes $V$ starts to 
decrease (see Figure~\ref{calib}). This behaviour, however, does not 
affect the diagnostic potential of the new line pairs as long as the 
line broadening is properly taken into account (which can easily be 
done by fitting the observed line profile).}

\end{appendix}

\end{document}